\documentclass[12pt,preprint]{aastex}

\shorttitle{Changes in solar structure}
\shortauthors{Basu and Mandel}

\begin{document}

\title{Does solar structure vary with solar magnetic activity?}

\author{Sarbani Basu and Anna Mandel}
\affil{Department of Astronomy, Yale University, P. O. Box 208101,
New Haven CT 06520-8101, U.S.A.}
\email{basu@astro.yale.edu, anna.mandel@yale.edu}

\begin{abstract}
We present evidence that solar structure changes with changes in solar activity.
We find that  the adiabatic index, $\Gamma_1$,
changes near  the second helium ionization, i.e., at a depth of about $0.98$R$_\odot$.
We believe that this change is a result of the change in the effective equation of
state caused by magnetic fields.
Inversions should be able to detect the changes in $\Gamma_1$  if mode sets 
with reliable and precise high-degree modes are available.
\end{abstract}


\section{Introduction}
\label{sec:intro}

It is well known that frequencies of solar oscillations vary with time
(Elsworth et al.~1990; Libbrecht \& Woodard 1990; 
Howe, Komm \& Hill 1999, etc.).
These variations are also known to be  correlated with solar activity.
However, these frequency changes appear to result from variations in
solar structure close to the surface, and no evidence has been found 
of structural changes deep in the solar interior (Basu 2002; Eff-Darwich et al. 2002;
Vorontsov 2001; 
Monteiro et al. 2000, etc.).
This is in contrast to solar dynamics, which shows changes correlated with
solar activity (Basu \& Antia 2003; Vorontsov et al. 2002; 
Howe et al. 2000; and references therein). 
Basu (2002) showed that the currently available mode sets do not allow
us to make any statement about changes in solar structure  above about $0.95$R$_\odot$ 
through inversions; the
lack of high degree modes makes the resolution of the inversions in these
layers  poor  and errors in the inversion results large.

\begin{figure}
\plotone{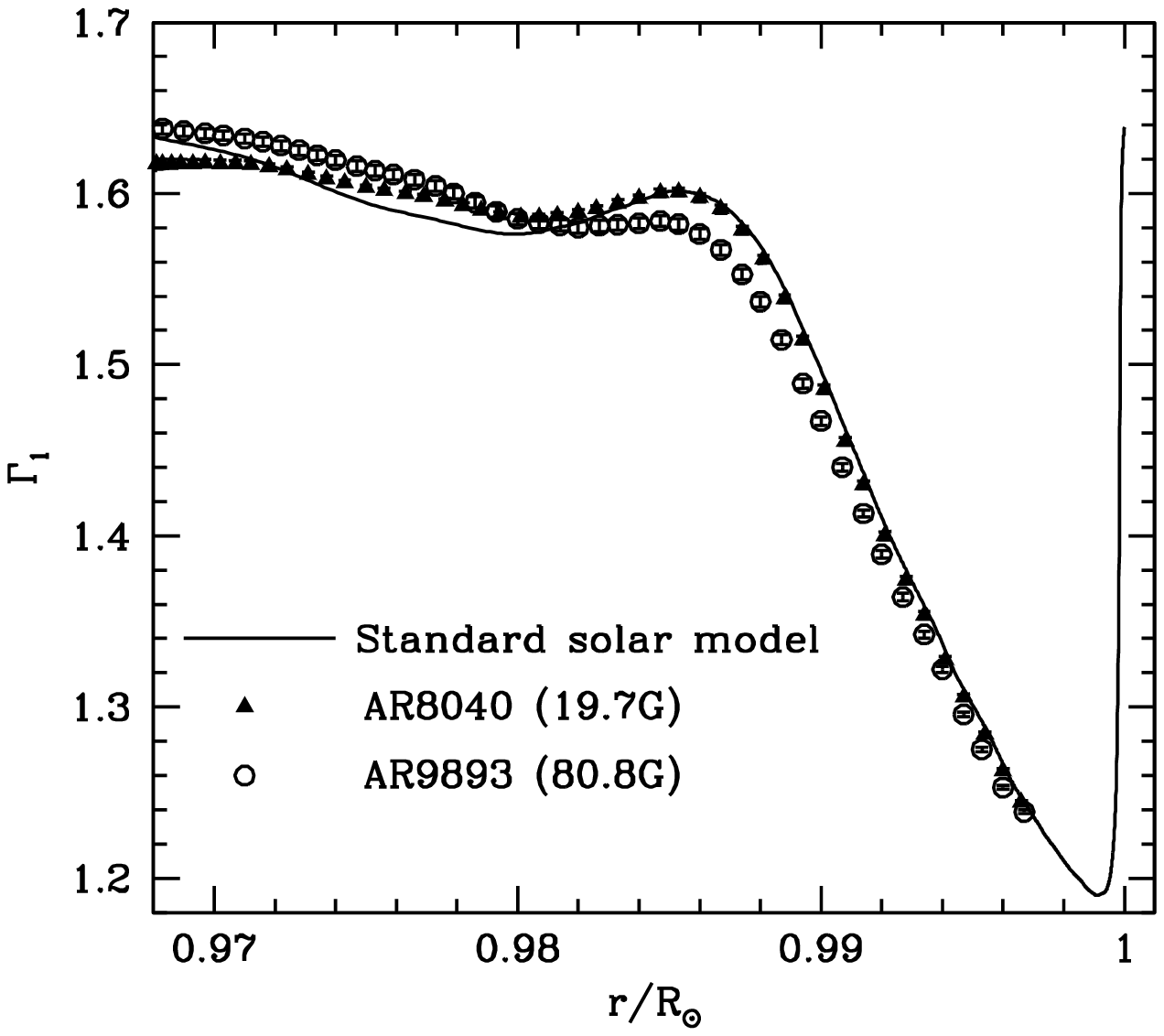}
\caption{The adiabatic index $\Gamma_1$ for a standard solar model and
two active regions. The region AR9893 is much stronger than AR 8040 and
their magnetic indices as defined in Basu et al. (2004) are shown in the figures.
The  $\Gamma_1$ profiles for the active regions were determined  from results 
of Basu et al. (2004) assuming that the
$\Gamma_1$ profile  of the quiet regions of the Sun is like that of a standard solar model.
The results would not change appreciably if, instead of a standard solar model, $\Gamma_1$
obtained from inversions of quiet-sun data had been used.
\label{fig:f1}}
\end{figure}

There have been recent studies of solar active regions using local helioseismology techniques
that suggest that active regions have lower sound speed compared with quiet regions
just below the surface 
(Basu, et al. 2004; Kosovichev et al. 2000, 2001).   It has also been shown 
by Basu et al. (2004) that the
adiabatic index, $\Gamma_1$ of active regions is considerably different from that of
quiet regions and the amount of change increases with increasing strength of
the active region. Figure~\ref{fig:f1} shows the adiabatic index for two of the active regions
studied by Basu et al. (2004), and one can see that the magnitude of the depression
in $\Gamma_1$ at  the second helium
ionization zone decreases with increasing strength of the active region.

This leads us to question whether similar changes occur globally in the Sun as
solar activity increases --- of course one would expect the changes to be much smaller
than those seen in active regions because global activity levels are much smaller
than in active regions. Although the He II ionization zone is too
shallow for studying temporal changes directly by inverting the currently available data 
sets, there
are indirect ways by  which we can study this region.

Any spherically symmetric localized, sharp feature or discontinuity in the
Sun's internal
structure leaves a definite signature on the solar p-mode frequencies.
Gough (1990) showed that abrupt changes of this type
contribute a characteristic oscillatory component to the frequencies
$\nu_{n,\ell}$ of those modes which penetrate below the
localized perturbation.
The amplitude of the oscillations increases with increasing
``severity'' of the discontinuity, and the wavelength of the oscillation
is essentially the acoustic depth of the sharp-feature.
Solar modes encounter two such features, the base of the convection zone (henceforth CZ) and
the He II ionization zone. The transition of the temperature gradient from the adiabatic to
radiative values at the CZ base
 gives rise to the oscillatory signal
in frequencies of all modes which penetrate below the 
CZ base; the depression in the adiabatic index $\Gamma_1$ in the 
He II ionization zone causes the second signal. The two oscillatory signals have very different
wavelengths and hence can be decoupled.
Not all modes see both features, only low degree modes ($\ell \la 25$) see
the CZ base, higher degree modes only see the ionization zone,
and the very high degree modes see neither. 
This signal has been used previously to study the
CZ base (Monteiro, Christensen-Dalsgaard \& Thompson 1994; 
Basu, Antia \& Narasimha 1994), and also study  changes in that region (Monteiro  et al. 2000). 

In this work we use the oscillatory signal from the depression in $\Gamma_1$ to study whether or not there
are changes in $\Gamma_1$ in that region that are correlated with solar activity.

\section{Technique}
\label{sec:tech}

The amplitude of the signals from the CZ base as well as from the
He II ionization zone are small, hence we first amplify them by taking the fourth differences
of the frequencies:
\begin{equation}
\delta^4\nu_{n,\ell}=\nu_{n+2,\ell}-4\nu_{n+1,\ell}+6\nu_{n,\ell}
-4\nu_{n-1,\ell}+\nu_{n-2,\ell}.
\label{eq:4th}
\end{equation}
Taking the fourth differences amplifies the signals  enough to be able to isolate
them, keeping the errors at a manageable level (see Basu, Antia \& Narasimha  1994 for a 
discussion).

The fourth differences  can
be fitted to the functional form given by Basu (1997):
\begin{eqnarray}
\delta^{4}\nu
&= \left[a_1+ a_2\nu +{(a_3+a_4L)/\nu^2}\right]\; +\nonumber\\
&  \left(a_5+{a_6\over\nu_m^2}
 +{a_7L \over\nu_m^2}+{a_8L\over\nu_m^4}\right)\sin(2\nu_m\tau_a
 +\psi_a)\; + \nonumber\\
&
\left(b_1+{b_2\over\nu_n^2}+{b_3L\over\nu_n^2}+{b_4L\over\nu_n^4}\right)
\sin(2\nu_n\tau_b+\psi_b),
\label{eq:func}
\end{eqnarray}
where
\begin{equation}
L=\ell(\ell+1), \quad
\nu_m=\nu-{{\gamma_aL}/{2\tau_a\nu}},\hbox{ and},\quad
\nu_n=\nu-{{\gamma_bL}/{2\tau_b\nu}}.
\label{eq:deff}
\end{equation}

 The coefficients $a_1$--$a_4$ define an overall smooth term,  $a_5$--$a_8$,
$\tau_a$ and $\psi_a$
define the oscillatory contribution due to the He II ionization zone and the remaining
terms define the oscillatory contribution from the CZ base.
The frequencies of the two components,
$\tau_a$ and $\tau_b$ are approximately the acoustic depths of the
He II ionization zone and the CZ base respectively,
but they also include a contribution from the frequency dependent part
of the phases $\psi_a$ and $\psi_b$ which is not taken into account explicitly.
The coefficients in this expression are determined by a least squares
fit to the fourth differences.

We use oscillation frequencies determined by the Michelson Doppler Imager (MDI) on
board the SOHO spacecraft, as well as data from the Global Oscillations Network Group (GONG).
We have used 38 data sets from MDI (Schou 1999), each covering
a period of 72 days starting from 1996 May 1 and ending on 2004 March 19.
We use 28  data sets from GONG
(Hill et al.~1996), each covering a period of
108 days, starting from 1995 May 7 and ending on 2003 August 16.  As a measure
of the solar activity for each data set, we use the mean radio flux at 10.7 cm during the 
time interval covered by each data set
as obtained from the US National Geophysical Data Center (www.ngdc.noaa.gov/stp/stp.html).

For each data set we use modes with two sets of degrees; one set consists of modes
with degree $5\le\ell\le25$ and we refer to this as the low-degree set, the second set
consists of all modes with $\ell\le60$  that have their lower 
turning point, $r_t > 0.715$ R$_\odot$ and we refer to this as the intermediate-degree set.
The lowest degree modes that satisfies this criterion for our frequency range of 2-3.5mHz
is $\ell=30$. Very few modes below $\ell$ of 42  satisfy this
criterion.
The first set of modes sample both the CZ base and the He II ionization zone and hence
we fit the entire  form given by Eq.~(\ref{eq:func}), while modes in the second set  sample only
the He II ionization zone and hence we fit the mode only to the first two terms of 
Eq.~(\ref{eq:func}). We do not use  modes with $\ell < 5$ because they have larger errors.
We do not use modes with degree higher than $\ell=60$ because the degree-dependence of
the signal becomes difficult to model and fit. 

\begin{figure}
\plotone{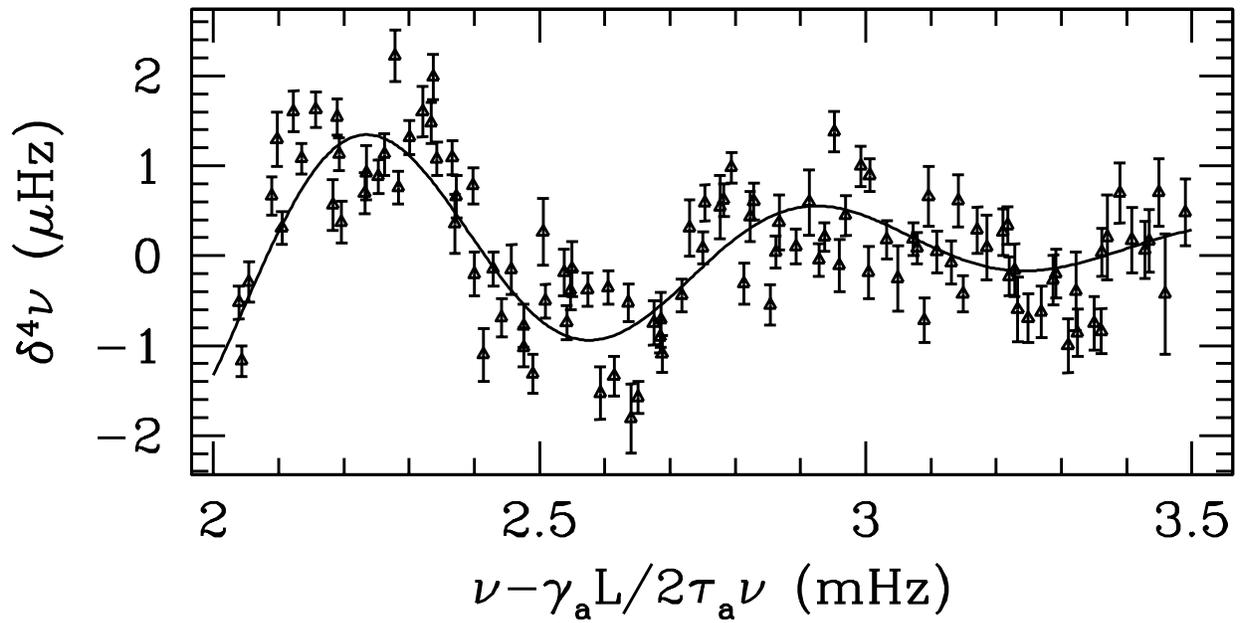}
\caption{The fourth-difference of frequencies of a set of GONG data with the degree-dependence
in the signal removed. Also shown is a fit to the signal in the $\Gamma_1$ depression.
Data for frequencies with $5\le\ell\le25$ are shown. 
\label{fig:fit4}}
\end{figure}

The quantity we are most interested in is the amplitude of the oscillatory signal from
He II ionization zone as a function of the magnetic activity level of the Sun when the data
were obtained.  Since the amplitude is both  degree-dependent as well as 
frequency-dependent, we use the averaged amplitude in the frequency range 2 to 3.5mHz
after the degree dependence is removed. This is the same as what was done by
Basu et al. (1994), Basu \& Antia (1994) and Basu (1997). The error on the result is
determined by Monte-Carlo simulations.
The fit to the signal from the He II ionization zone, after removal of the
degree dependence,   is shown 
in Fig.~\ref{fig:fit4} for one set of low-degree GONG data.

\section{Results}
\label{sec:res}

\begin{figure}
\plotone{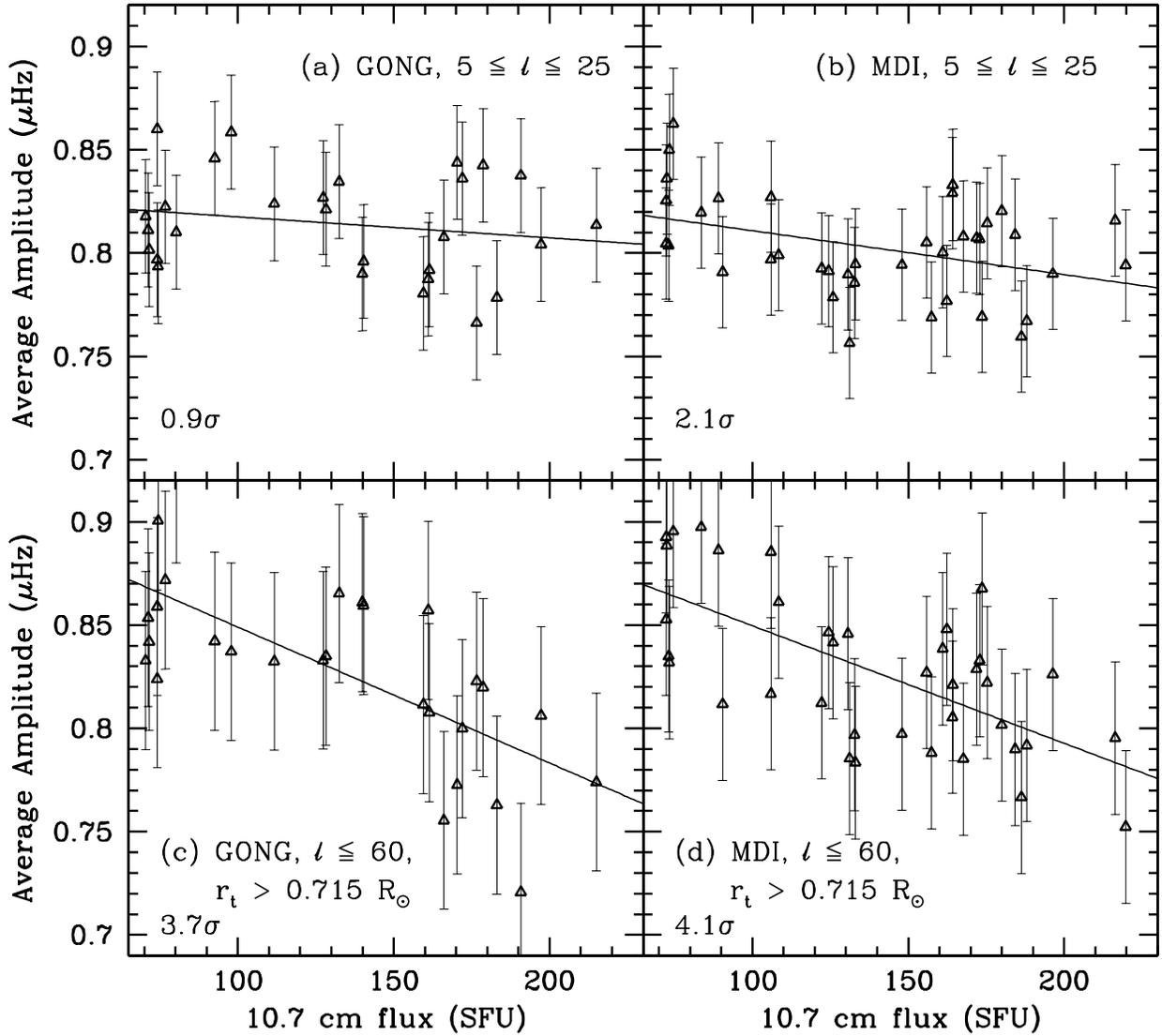}
\caption{The amplitude of the oscillatory signal averaged over the frequency
range 2-3.5 mHz plotted as a function of the 10.7 cm radio flux, which is an
indicator of solar activity. The 10.7 cm flux is plotted in units of 
$10^{-22}$ J s$^{-1}$ m$^{-2}$ Hz${-1}$ (Solar flux units).  The different panels show
the results of the different degree ranges of the GONG and MDI sets. The
linear least squares fit to the points is  shown as a continuous line. Each panel also lists
how many standard deviations away the slope of the fitted line is away from zero.
\label{fig:sepa}}
\end{figure}

The amplitudes of the oscillatory signal that arises from the He II ionization zone 
are plotted as a function of the solar activity index in Fig.~\ref{fig:sepa}. The
10.7 cm flux is in units of the Solar Flux Unit (SFU), i.e., $10^{-22}$ J s$^{-1}$
m$^{-2}$ Hz$^{-1}$. We have plotted the four sets of data separately.

We find that in all cases, the amplitude decreases with increasing solar activity,
a result that is expected if the active-region results can be applied to the global
Sun. We can fit straight lines to the data quite easily --- the large errors do not
justify fitting more complicated trends. We can see that for all 4 sets, the straight 
line has a finite slope, however, the slope is not always statistically significant.
This is particularly true for the GONG low-degree data. The MDI low-degree sets show only
a marginally significant decrease. However, the intermediate-degree sets for both GONG
and MDI show a reasonably significant trend ($\approx  4\sigma$). The scatter in the plots is
somewhat less than  what the errors on the points should suggest, so it is likely that the 
errors have been overestimated and significance of the slope  underestimated.

It is not completely surprising that the low- and intermediate-degree sets give us
somewhat different results. For the low-degree sets, we need to correctly remove the
signal from the CZ base to get the correct amplitude from the
He II ionization zone. 
Simulations performed with different solar models show that
this process leads to substantial systematic errors in the results. 
The intermediate-degree modes are not affected by the CZ base at all, and hence
the signal due to the He II zone is cleaner and easier to measure. Thus we
put more weight on the results obtained from the intermediate degree modes.

\begin{figure}
\plotone{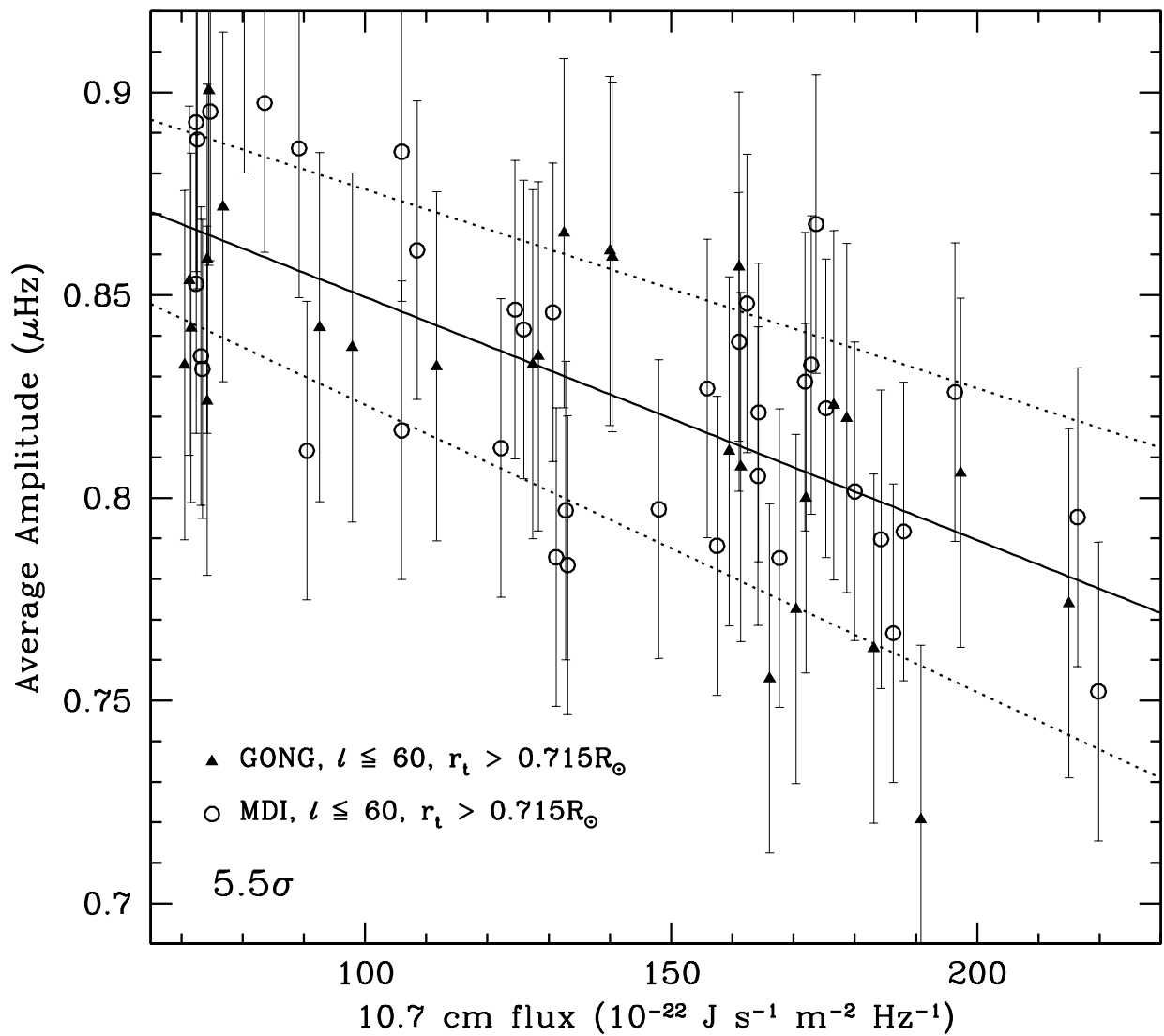}
\caption{The GONG and MDI results obtained for modes with  $\ell\le 60$ and $r_t > 0.715$R$_\odot$
 plotted as a function of the 10.7 cm flux. The black continuous line is a least-squares
fit to all the data points. The dotted lines show the $1\sigma$ spread because of
errors in the fitted parameters.
\label{fig:high}}
\end{figure}

Figure~\ref{fig:high} shows both GONG and MDI intermediate-degree results plotted as 
a function of the 10.7 cm flux. It is clear that both GONG and MDI show very similar
trends, which is encouraging since these are independent projects. A straight line fit
to all the points shows that the slope is 5.5$\sigma$ from zero, a result that is
reasonably statistically significant. 
 We therefore, can conclude with a degree of confidence that
the region of the He II ionization zone changes with change in solar activity. In particular,
the magnitude of the dip in $\Gamma_1$, which is what causes the signal we are looking at,
decreases with increasing activity.
It should be noted however, that there should be a correlated noise component between
the GONG and MDI data sets since they are observing the same object, hence the
increase in the significance may be smaller than what we find on combining the results.
A combination of all the results (GONG and MDI, low and intermediate-degree set) gives
a slope significant at the 4.7$\sigma$ level.

\section{Discussion}
\label{sec:disc}

The easiest interpretation of a change in the magnitude of the depression of $\Gamma_1$
 at the He II ionization zone is
a change in the abundance of helium. However, that interpretation does not apply in this
case, since we are talking of cyclical changes over very short time-scales. The only 
change in helium that is expected in the CZ is a monotonic decrease due to the gravitational 
settling of helium that takes place over  very long time-scales. We need to look at the
the equation of state (EOS) to understand the changes. The presence of magnetic
fields can change the effective EOS because of contributions of the magnetic fields 
to  energy and pressure. 
A change in the EOS 
changes the shape and magnitude of the dip in $\Gamma_1$ at the He II ionization zone for 
the same helium abundance.

\begin{figure}
\plotone{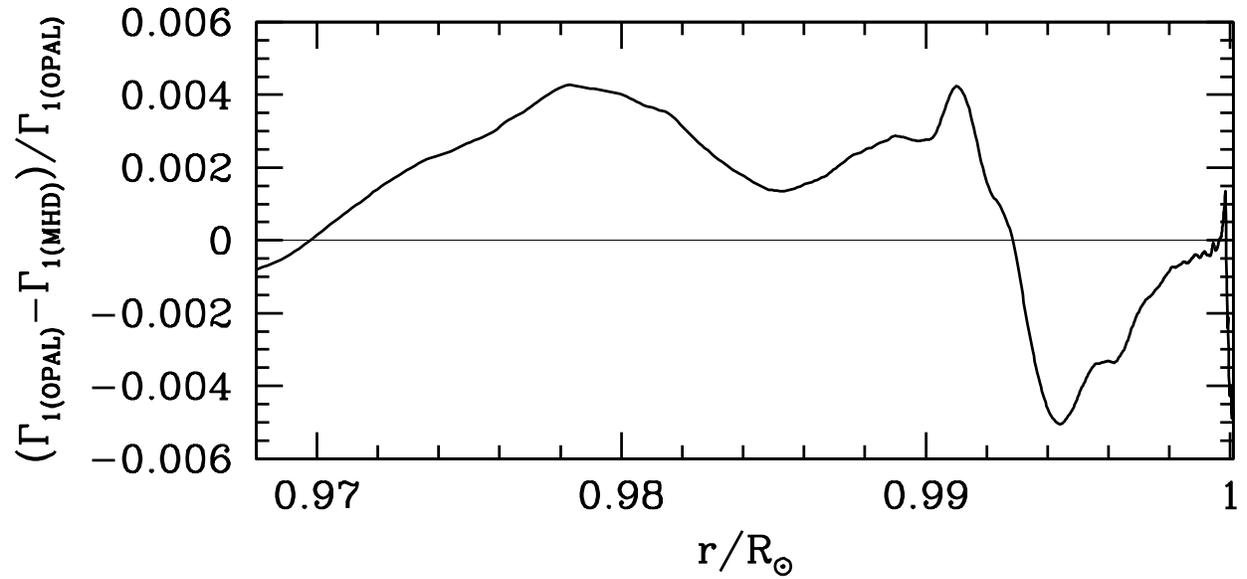}
\caption{The relative difference in the adiabatic index $\Gamma_1$ between two
solar envelope models, one constructed with the MHD and the other with the OPAL
equation of state. The models were constructed with the same helium and heavy metal
abundances and the same CZ depth.
\label{fig:gamdif}}
\end{figure}

Figure~\ref{fig:gamdif} shows the relative difference in $\Gamma_1$ between two solar envelope
models, one constructed with the so-called MHD equation of state 
(Hummer \& Mihalas 1988; Mihalas, D\"appen \& Hummer 1988; D\"appen et al.~1988),
and the second with the OPAL equation of state (Rogers, Swenson \& Iglesias 1996). Both models
were constructed with the same opacities, have identical helium and heavy element abundances 
in the CZ (0.242 and 0.018 respectively) and were constructed to have the same CZ depth
(0.287R$_\odot$). We can see that the models have substantial differences in $\Gamma_1$
in the region of the He II ionization zone and higher. 
The amplitude of the
He II signal for the MHD model is 1.164$\mu$Hz and that of the OPAL model is 
1.027$\mu$Hz. 

The difference between the He II amplitudes of the two models (0.137$\mu$Hz) 
is larger than the total range of change 
in amplitude seen in Fig.~\ref{fig:high} (which is only $0.099\mu$Hz according to
the linear fits to the results). Thus as a first approximation we can say that the 
change in $\Gamma_1$ near the solar helium ionization zone between solar minimum and
maximum is less than that in Fig.~\ref{fig:gamdif}, i.e., less than about 4\%. However, one
must be careful about this number; the $\Gamma_1$ differences in Fig.~\ref{fig:gamdif} 
have some fairly sharp features that could be the cause of the large difference in amplitude
for the two models. It is unlikely that a small change in the EOS due to
magnetic fields would cause spiky changes, and hence for  $0.099\mu$Hz could imply a 
larger difference in $\Gamma_1$. 
Inversions  could easily detect differences in $\Gamma_1$ less than the estimated
$\Gamma_1$ change between solar minimum and solar maximum. However, to do so reliably,
and with any degree of statistical significance, we need mode sets with reliable and
precise high-degree modes (Di Mauro et al. 2002; Rabello-Soares et al. 2000).

\begin{figure}
\plotone{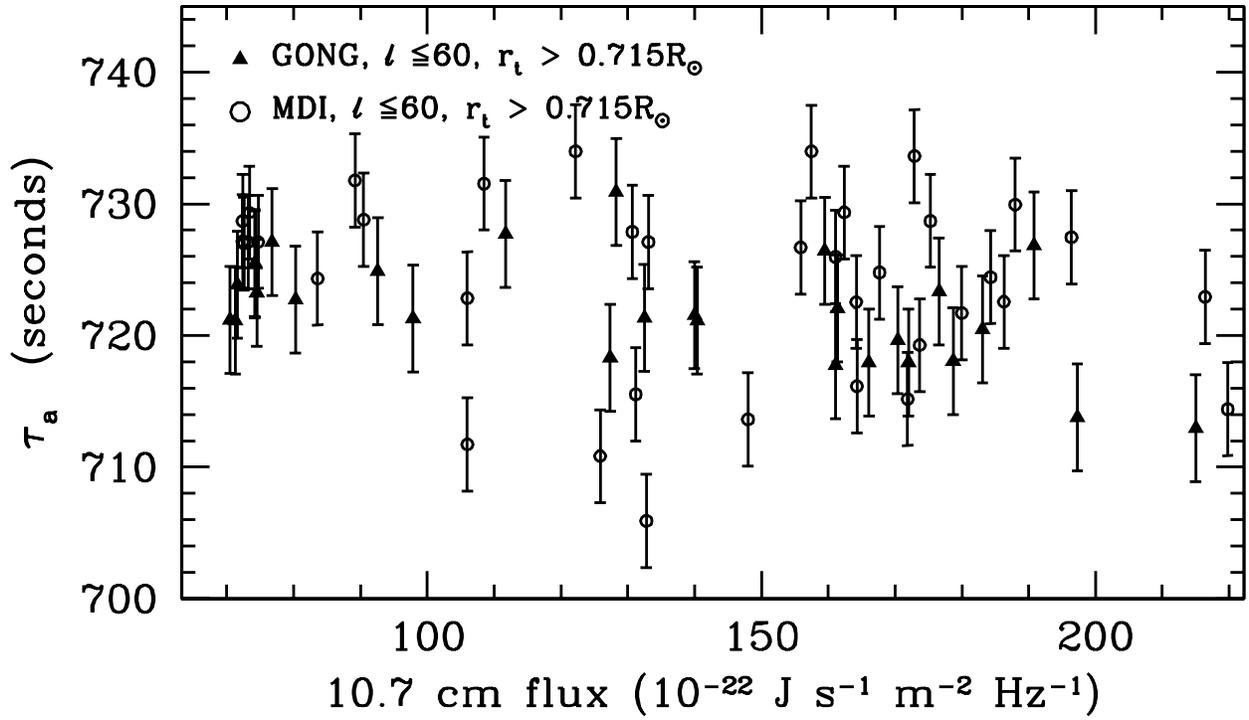}
\caption{The variation of $\tau_a$ as a function of the 10.7cm flux. 
\label{fig:tau}}
\end{figure}

As far as the parameter $\tau_a$ is concerned, we cannot draw any conclusions since
as can be seen 
 in Fig.~\ref{fig:tau}, the points are widely scattered.

One expects solar cycle related changes to have strong latitudinal dependences. We repeated the
study described above  for different latitudes, but the errors are too large to be able to 
observe any latitudinal dependence.

\section{Conclusions}
\label{sec:conclu}

We find evidence that  suggests that solar structure changes with change in
solar activity in the layers around the He II ionization zone (i.e., 0.98 R$_\odot$
and thereabouts). The depression in the adiabatic index $\Gamma_1$ in the
He II ionization zone decreases with increasing solar activity. 
This is the first
evidence to suggest that solar structure changes with solar activity. 

\acknowledgments

This work  utilizes data obtained by the Global Oscillation
Network Group project, managed by the National Solar Observatory
which is
operated by AURA, Inc. under a cooperative agreement with the
NSF.
This work also utilizes data from the Solar Oscillations
Investigation/ Michelson Doppler Imager (SOI/MDI) on the Solar
and Heliospheric Observatory (SOHO).  SOHO is a project of
international cooperation between ESA and NASA.
The MDI project is supported by NASA contract NAG5-13261 to Stanford 
University.
This work was supported by
NSF grants ATM 0206130 and ATM 0348837 to SB

\clearpage

\clearpage

\clearpage

\clearpage

\clearpage

\clearpage

\begin{thebibliography}{}


\bibitem[Basu(1997)]{ba97}
Basu, S. 1997, MNRAS 288, 572

\bibitem[Basu(2002)]{ba02}
Basu, S. 2002, in From Solar Min to Max: Half a Solar Cycle with SOHO,
Proc. SOHO 11 Symposium, ed. A. Wilson, ESA SP-508, 7

\bibitem[Basu \& Antia(1994)]{bahma94}
Basu, S., Antia, H. M. 1994, MNRAS 269, 1137


\bibitem[Basu \& Antia(2000)]{bahma00b}
Basu, S., Antia, H. M. 2000, ApJ, 541, 442

\bibitem[Basu \& Antia(2003)]{bahma03}
Basu, S., Antia, H. M., ApJ,  2003, 585, 553

\bibitem[Basu et al.(1994)]{baet94}
Basu, S., Antia, H. M., Narasimha, D. 1994, MNRAS 267,209

\bibitem[Basu et al.(2004)]{baet04}
Basu, S., Antia, H. M., Bogart, R. S. 2004, ApJ, 610. 1157


\bibitem[D\"appen et al.(1988)]{da88}
D\"appen W., Mihalas D., Hummer D. G., Mihalas B. W. 1988, ApJ  332, 261

\bibitem[Di Mauro et al.(2002)]{dimau02}
Di Mauro, M. P., Christensen-Dalsgaard, J., Rabello-Soares, M. C. and
Basu, S. 2002, A\&A 384, 666


\bibitem[Eff-Darwich et al.(2002)]{ef02}
Eff-Darwich, A., Korzennik, S. G., Jiménez-Reyes, S. J., Pérez Hernández
2002, ApJ, 580, 574

\bibitem[Elsworth et al.(1990)]{els90}
Elsworth, Y., Howe, R., Isaak, G. R., McLeod, C. P., 
New, R. 1990,  Nature, 345, 322

\bibitem[Gough(1990)]{gou90}
Gough, D. O. 1990, in Lecture Notes in Physics, eds., Y. Osaki, H. Shibahashi, 
(Springer: Berlin), vol. 367, p 283.

\bibitem[Hill et al.(1996)]{hill96}
Hill F., et al.~1996, Science 272, 1292

\bibitem[Howe et al.(1999)]{ho99}
Howe, R., Komm, R.,  Hill, F. 1999,  ApJ, 524, 1084

\bibitem[Howe et al.(2000)]{ho00}
Howe, R., Christensen-Dalsgaard, J., Hill, F., Komm, R. W., Larsen, R. M., 
Schou, J., Thompson, M. J., Toomre, J. 2000, ApJ, 533, L163

\bibitem[Hummer \& Mihalas(1988)]{hu88}
Hummer D. G., Mihalas D. 1988,  ApJ  331, 794

\bibitem[Libbrecht \& Woodard(1990)]{li90}
Libbrecht, K. G.,  Woodard, M. F. 1990,  Nature, 345, 779

\bibitem[Miglio et al.(2003)]{mi03}
Miglio, A., Christensen-Dalsgaard, J., di Mauro, M. P., Monteiro, M. J. P. F. G., Thompson, M. J.
2003, in Asteroseismology Across the HR Diagram, eds. M.J. Thompson, M.S. Cunha, 
M.J.P.F.G Monteiro (Kluwer:Dordrecht)  p. 537

\bibitem[Mihalas et al.(1988)]{mi88}
 Mihalas D., D\"appen W., Hummer D. G. 1988,  ApJ  331, 815

\bibitem[Monteiro et a.(1994)]{mo94}
Monteiro, M. J. P. F. G., Christensen-Dalsgaard, J., Thompson, M. J.
1994, A\&A, 283, 247

\bibitem[Monteiro et al.(2000)]{moet00}
Monteiro, M. J. P. F. G., Christensen-Dalsgaard, J., Schou, J., Thompson, M. J.
2000, in  Helio- and asteroseismology at the dawn of the millennium, Proc. SOHO 10/GONG 2000 Workshop,
ed. A. Wilson,  ESA SP-464, 535

\bibitem[Kosovichev et al.(2000)]{kos00}
Kosovichev, A. G., Duvall, T. L., Jr., Scherrer, P. H. 2000,
Solar Phys., 192, 159

\bibitem[Kosovichev et al.(2001)]{kos01}
Kosovichev, A. G., Duvall, T. L., Jr., Birch, A. C., Gizon, L.,
Scherrer, P. H., Zhao, J. 2001, in Helio- and Asteroseismology at
the Dawn of the Millennium, Proc. SOHO 10/GONG 2000 Workshop,
ed., A. Wilson, ESA SP-464, 701


\bibitem[Rabello-Soares et al.(2000)]{ra00}
Rabello-Soares, M. C., Basu, S., Christensen-Dalsgaard, J., Di Mauro, M. P. 2000,
Solar Phys. 193, 345

\bibitem[Rogers et al.(1996)]{ro96}
Rogers, F. J., Swenson, F. J., Iglesias, C. A. 1996, ApJ 456, 902


\bibitem[Schou(1999)]{sch99}
Schou J., 1999, ApJ 523, L181

\bibitem[Vorontsov(2001)]{vo01}
Vorontsov, S. V. 2001, in Helio- and Astero-seismology at the dawn of the millennium: 
Proc. SOHO 10/GONG 2000 Workshop,  ed. A. Wilson, ESA SP-464,  563

\bibitem[Vorontsov et al.(2002)]{vo02}
Vorontsov, S. V., Christensen-Dalsgaard, J., Schou, J., Strakhov, V. N., Thompson, M. J.
2002, Science, 296. 101

\end{thebibliography}
\end{document}